\documentclass[aps,amsmath,showpacs,showkeys,superscriptaddress,twocolumn]{revtex4}
\usepackage{color}
\usepackage{graphicx}
\usepackage{amssymb}
\usepackage{subfigure}
\usepackage{titlesec}
\usepackage{amssymb,amsfonts,amsthm}

\begin{document}
\title{Travelling-wave solutions and solitons of KdV, mKdV and NLS equations}
\author{Supriya Chatterjee}
\affiliation{Department of Physics, A. P. C. Roy Government College, Himanchal Vihar, Matigara, Siliguri, 734010, West Bengal, India}
\email{supriyo.thiba@gmail.com}
\author{Pranab Sarkar}
\affiliation{Department of Chemistry, Visva-Bharati University, Santiniketan 731235, India}
\author{Benoy Talukdar}
\affiliation{Department of Physics, Visva-Bharati University, Santiniketan 731235, India}
\begin{abstract}
We introduce the concept of soliton solutions of integrable nonlinear partial differential equations and point out that the inverse spectral method represents the rigorous mathematical formalism to construct such solutions.  We work with the travelling waves of the KdV, mKdV and nonlinear Schr\"{o}dinger (NLS) equations and derive a pedagogic method to find their soliton solutions. The travelling wave of the KdV equation leads directly to the well known bell type KdV soliton while the mKdV equation needs some additional consideration in respect of this. The travelling waves of a generalized mKdV and NLS equations are obtained in terms of $sn(u,m)$, the so called Jacobi  elliptic sine function. The choice $m=1$ provides a constraint on the parameters of the equations and gives their kink and anti-kink soliton solutions. We further show that by expressing the travelling waves of these equations in terms of Jacobi  elliptic cosine functions, $cn(u,m)$, it is possible to construct bell-type soliton solutions.                                                                                                              
\end{abstract}
\maketitle
\section*{1. Introduction}
Sir Isaac Newton taught us how to model motion of material bodies in terms of equations. Subsequently, generations of physicists and mathematicians extended the idea and used differential equations to study time evolution of physical processes. A differential equation expresses a relation between derivatives of a given function  of certain variables with the function itself and thus provides a mathematical procedure to  describe the law of nature in terms of small differences of time and space. Differential equations are called ordinary if there is only one independent variable, and partial if there are more than one. In an ordinary differential equation only ordinary derivatives appear, while a partial differential equation contains partial derivatives. Both ordinary and partial differential equations may either be linear or nonlinear.  A linear differential  equation is one in which  there appears only the first power of the function or derivatives of the function. On the other hand, a nonlinear differential equation involves higher powers (equal to or greater than two) of the function or its derivatives. 
\par There are standard analytic approaches to solve linear differential equations \cite{1,2}. This is, however, not true if the equations are nonlinear \cite{3}. But there are few nonlinear differential equations which can be solved by analytic methods. These equations are said to form what is referred to in the literature as integrable system \cite{4}. The equations of the integrable system supports soliton solutions. Solitons are solitary waves of constant shape. These localized waves are stable against mutual collision \cite{5}.
\par The wave of unchanging shape was first observed in 1834 by a Scottish engineer, John Scott Russell \cite{6}. Soliton was, however, discovered much later by Zabusky and Kruskal \cite{7} through a computer experiment to obtain solutions of the periodic Korteweg-de Vries (KdV) equation. In the standard notation the KdV equation is given by
\begin{equation}
 V_t+6VV_x+V_{3x}=0,\;\;\;\;\;V=V(x,t).
\end{equation}
Here $x$ and $t$ stand for space and time variables such that Eq. (1) represents a wave propagation in the $x$ direction. The subscripts of $V$ denote appropriate partial derivatives and Eq. (1) stands for a nonlinear partial differential equation.  The linear term $V_{3x}$ accounts for the dispersion of the wave while the nonlinear term takes care of its squeezing due to interaction with the medium. A delicate balance between the dispersive and squeezing effects in the medium produces the so-called soliton. 
\par There exists a generalization of Eq. (1) written as \cite{8} 
\begin{equation}
 \phi_t+6\phi^2\phi_x+\phi_{3x}=0,\;\;\;\;\phi=\phi(x,t).
\end{equation}
Equation (2) is often called the modified KdV or mKdV equation. The solutions of KdV and mKdV equations are related by a nonlinear transformation
\begin{equation}
 V=\phi^2+\phi_x
\end{equation}
discovered by Miura [9]. Equations (1) and (2) differ only in the degree of nonlinearity. Interestingly, the transformation in Eq. (3) is a Riccati equation \cite{3} so that it can be reduced to the linear form
\begin{equation}
 \psi_{xx}-V\psi=0
\end{equation}
by the use of Cole-Hopf transformation \cite{10,11} written as
\begin{equation}
 \phi=\frac{\psi_x}{\psi}.
\end{equation}
\par In 1967 an ingenious method was developed by Gardner, Greene, Kruskal and Miura (GGKM) \cite{12} to solve the KdV equation analytically. In this approach $V(x,t)$ is determined from the solution of a Gel'fand-Levitan-Marchenko equation \cite{13} using the spectral data of a one-dimensional Schr\"{o}dinger equation for a potential $V(x,0)$ as input. The formalism developed in ref. 12 goes by the name inverse spectral transform method. The seminal work of GGKM raised KdV equation to the status of first member of the so-called integrable system. Soon afterwards, in 1972, Wadati \cite{14} showed that mKdV equation can also be solved with the help of inverse scattering problem associated with a coupled first-order linear differential equation.
\par The KdV equation governs small but finite waves in shallow water \cite{15}. On the other hand, the mKdV equation
arises in plasma physics and lattice dynamics \cite{16}. Next to KdV and mKdV equations there is another nonlinear evolution equation, the so-called nonlinear Schr\"{o}dinger equation. This equation appears in many applicative contexts \cite{17}. In general, it is appropriate to describe the time evolution of the envelope of an almost monochromatic wave of moderate amplitude in a weakly nonlinear dispersive system. The nonlinear Schr\"{o}dinger (NLS) equation is given by
\begin{equation}
 iu_t+u_{xx}+\lambda|u|^2u=0,
\end{equation}
with, $u=u(x,t)$, a complex function and $\lambda$, a real constant. In a remarkable paper Zakharov and Shabat \cite{18} reported the inverse spectral problem needed to solve Eq. (6) analytically. As opposed to the case of KdV equation one needs a matrix Schr\"{o}dinger equation \cite{19} to obtain the solution of Eq.(6) and thus a useful result for the so-called optical soliton. 
\par In the present work we shall make use of the travelling wave solution of the KdV, mKdV and NLS equations to obtain their soliton solutions. Here our objective is to provide an uncomplicated mathematical approach to compute soliton solutions of these equations. Being invariant under translation the equations of our interest will reduce to ordinary nonlinear differential equations in the travelling coordinate. We shall see in the course of our study that the reduced equations so obtained admit simple analytic solutions. 
\par The method to find travelling wave solution of almost all equations proceeds by a simple substitution which transforms any given partial differential equation for $\omega(x,t)$ to an ordinary one. In (1+1) dimensions, the simplest mathematical form of a travelling wave is a function $f(\xi)$  related to $\omega(x,t)$ by
\begin{equation}
 \omega(x,t)=f(\xi),\;\;\;\;\xi=x-ct.
\end{equation}
Here $c$ stands for the velocity of the wave such that $ct$ has the dimension of length. The quantity $\xi$ is often referred to as the travelling coordinate. At $t=0$ the wave has the form $f(x)$ which is the initial wave profile. Then $f(x-ct)$ represents the profile at a time $t$ that is just the initial profile translated to the right by spatial $ct$ units. It is straightforward to verify that $f(\xi)$ is a solution of the constant coefficient transport equation $\omega_t+c\omega_x=0$.
\par In section 2 we show that the travelling-wave solution of the KdV equation (1) leads, rather naturally, to the  well- known bell shaped non-topological soliton. Here we also deal with the mKdV equation (2) and make some useful comments on its travelling wave solution which bears a structural similarity  with that of the KdV equation. In addition to bell-shaped solitons there can be topological kink and anti-kink solitons \cite{20} for mKdV  and NLS equations. In view of this we provide here a brief introduction to such topological solitons with attention to the well known Sine-Gordon equation written as \cite{8}
\begin{equation}
 U_{xx}-U_{tt}=\sin U,
\end{equation}
where $U(x,t)$ is defined modulo $\pi$ subject to the boundary condition $U(x,t)\rightarrow0$ as $x\rightarrow\pm\infty$ for every finite $t$. We devote section 3 to deal with a generalized mKdV equation represented by
\begin{equation}
 \phi_t+\alpha\phi^2\phi_x+\beta\phi_{xxx}=0
\end{equation}
and the NLS equation (6). In equation (9) $\alpha$ and $\beta$ stand for arbitrary real constants. We obtain their travelling wave solutions and impose constraints on them to look for the corresponding soliton solutions. Finally, in section 4 we summarize our outlook on the present work and make some concluding remarks.
\section*{2. Soliton solutions of KdV, mKdV and sine-Gordon equations}
The travelling wave $V(x,t)=f(\xi)$ converts the KdV equation (1) in the form
\begin{equation}
 -cf'(\xi)+6f(\xi)f'(\xi)+f'''(\xi)=0
\end{equation}
which on integration gives
\begin{equation}
 f''(\xi)+3f(\xi)^2-cf(\xi)=0.
\end{equation}
In writing Eq. (11) we assumed that the boundary conditions of $f(\xi)$ and its derivatives with respect to $\xi$ are such that the constant of integration is zero. Multiplying Eq. (11) by $f'(\xi)$ we can integrate the resulting expression to get
\begin{equation}
 -\frac{2}{\sqrt{c}}\arctan h[\frac{\sqrt{c-2f(\xi)}}{\sqrt{c}}]=\xi
\end{equation}
such that
\begin{equation}
 f(\xi)=\frac{c}{2}\sec h^2[\frac{\sqrt c}{2}\xi].
\end{equation}
Clearly, the travelling wave solution in Eq. (13) represents the well-known soliton \cite{4} solution of the KdV equation.
\par We now proceed to find a similar travelling wave solution of the mKdV equation (2). To that end we introduce
\begin{equation}
 \phi(x,t)=f_m(\xi).
\end{equation}
The subscript $m$ of $f$ in Eq. (14) is self explanatory. Making use of Eq. (14) in Eq. (2) one can follow the treatment for the KdV equation to get the travelling wave solution of the mKdV equation in the form
\begin{equation}
 f_m(\xi)=\frac{2c\exp[\xi\sqrt{c}]}{c+\exp[2\xi\sqrt{c}]}.
\end{equation}
It is of interest to note that for $c=1$, Eq. (15) reads
\begin{equation}
 f_m(\xi)=\sec h\xi
\end{equation}
which represents  the fundamental soliton of the mKdV as well as NLS equation \cite{17}.
\par Let us compare the plots of $f(\xi)$ and $f_m(\xi)$ as a function of $\xi$ for values of $c$ less than and equal to one. In figures 1(a), 1(b) and 1(c) we portray the results of $f(\xi)$ and $f_m(\xi)$ and for $c=0.2$, 0.6 and 1.0 respectively. The solid curves give the variation of $f(\xi)$, the solution of the KdV equation while the dashed curves give similar variation of $f_m(\xi)$, the travelling wave of the mKdV equation.
\begin{figure}[htb!]
\includegraphics[width=4.5cm]{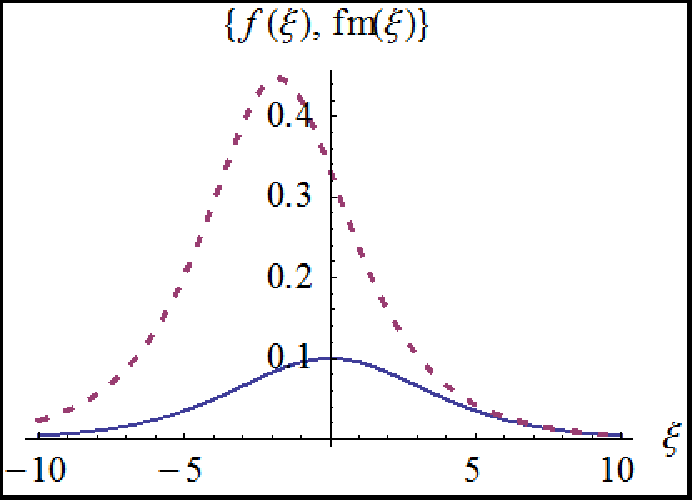}
\includegraphics[width=4.5cm]{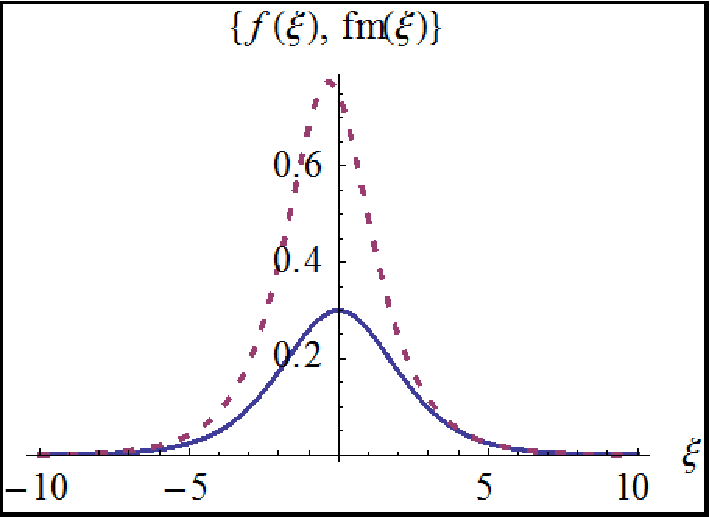}
\includegraphics[width=4.5cm]{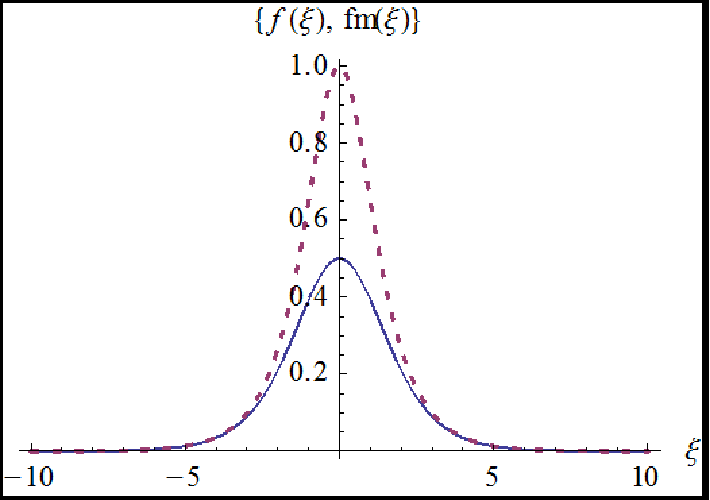}
\caption{(a) (upper) Travelling waves of KdV and mKdV for $c=0.2$, (b)(middle) Same as Fig. 1(a)but for $c=0.6$, (c) (lower) Same as Fig. 1(a)but for $c=1$.}
\end{figure}
In each of these figures the curve for $f(\xi)$ is symmetrical about the $\xi$ axis. This is quite expected since the travelling wave solution of the KdV equation is a soliton. Looking closely into these figures we see that height of the KdV soliton increases with increasing values of $c$. This is also true for the travelling wave of the mKdV equation. But as opposed to the travelling wave solution of the KdV equation, that for the mKdV equation is asymmetrical about $f_m(\xi)$ and/or $f(\xi)$ axis for $c<1$. Amongst the results presented the dashed curve in Fig.1(a) for $c=0.2$ exhibits maximum asymmetry. The dashed curve in  Fig.1(b) for $c=0.6$ appears to be less asymmetric compared  to  that in Fig.1(a). From Fig.1(c) we see that the curve for $f_m(\xi)$ for $c=1$ is symmetric about the $\xi$ axis. This is quite expected since for $c=1$ the travelling wave solution of the mKdV equation is actually the fundamental soliton solution of the mKdv and NLS equations. From Fig.1(c) we see that the height of the soliton formed by the travelling wave solution of the mKdV equation  for $c=1$ is bigger than that of the KdV soliton. The reason for this may be attributed to the simple fact that mKdV equation is more nonlinear than the KdV equation.
\par We now seek a travelling wave solution of the sine-Gordon equation (8). Making use of $U(x,t)=G(\xi)$ in Eq. (8) we get
\begin{equation}
 G''(\xi)=\gamma^2\sin G(\xi),\;\;\;\;\gamma=(1-c^2)^{-\frac{1}{2}}.
\end{equation}
We now multiply Eq.(17) by $G'(\xi)$ and integrate the resulting equation with respect to $\xi$ to get
\begin{equation}
 \frac{1}{2}(G'(\xi))^2=A-\gamma^2\cos G(\xi),
\end{equation}
where $A$ is a constant of integration. The boundary condition used for solving Eq. (8) is given by $U_x(x,t)\rightarrow\pm\infty$ as $x\rightarrow\pm\infty$. This gives $A=\gamma^2$ such that eq.(18) can be solved to get
\begin{equation}
 G(\xi)=4\arctan(e^{\pm\gamma\xi}).
\end{equation}
The solution in Eq. (19) goes by the name Kink (+sign) and Anti-kink (- sign) solitons \cite{20}. Figure 2 gives the plot of Kink,$G_+(\xi)$ , and Anti-kink, $G_-(\xi)$, solitons as a function of $\xi$ for $\gamma=0.4$.
\begin{figure}[htb!]
\includegraphics[width=4.5cm]{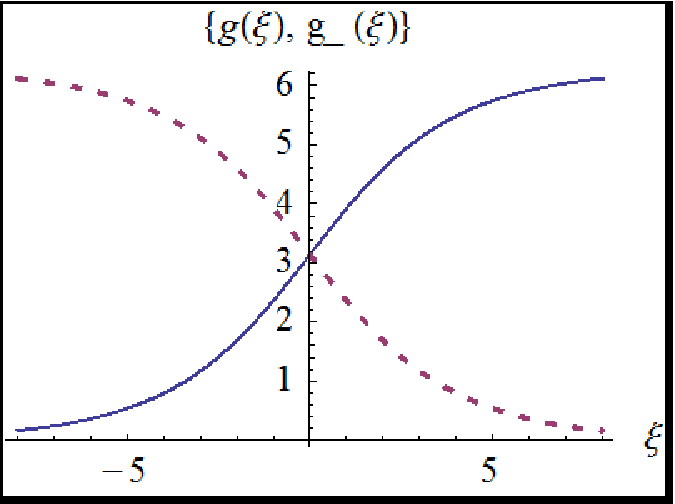}
\caption{The Kink and Anti-kink solitons for $\gamma=0.4$.}
\end{figure}
The solid and dashed curves show the plots for the Kink and Anti-kink solitons as a function of the travelling coordinate $\xi=x-ct$. In Figs. 3(a) and 3(b) we present 3D plots of the above Kink and Anti-Kink solitons for $c=1$.  
\begin{figure}[htb!]
\includegraphics[width=4.5cm]{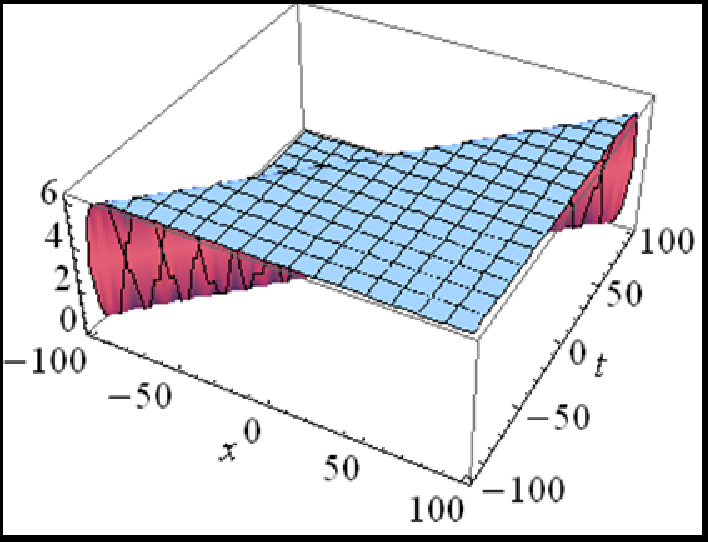}
\includegraphics[width=4.5cm]{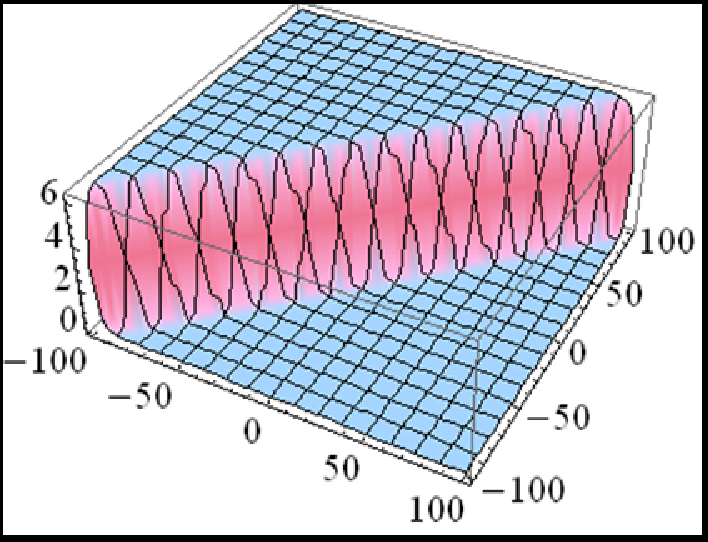}
\caption{(a) (upper) 3D plot of the Kink soliton, (b)(lower) 3D Plot of Anti-kink soliton.}
\end{figure}
\section*{3. Travelling waves and solitons of generalized mKdV and NLS equations}
Let $f_g(\xi)$ represent the travelling wave solution of the generalized mKdV equation (9) such that 
\begin{equation}
 -cf'_g(\xi)+\alpha f_g^2(\xi)f'_g(\xi)+\beta f'''_g(\xi)=0.
\end{equation}
Taking anti-derivative of Eq. (20) with zero integral constant we get 
\begin{equation}
 f''_g(\xi)-af_g(\xi)+bf^3_g(\xi)=0
\end{equation}
with
\begin{equation}
 a=\frac{c}{\beta}\;\;\;\;\mbox{and}\;\;\;\;b=\frac{\alpha}{3\beta}.
\end{equation}
Thus we see that the travelling wave solution of the generalized mKdV equation satisfies the differential equation for a non-dissipative Duffing oscillator \cite{21}.  Equation (21) can be integrated to get
\begin{equation}
 s-\xi=0,
\end{equation}
where
\begin{equation}
 s=-i\sqrt{\frac{2}{\delta_-}}F(i\arcsin h(-\frac{b}{\delta_+}v,m)),\;\;\;\;v=f_g(\xi)
\end{equation}
with
\begin{equation}
 m=\frac{\delta_+}{\delta_-}\;\;\;\;\mbox{and}\;\;\;\;\delta_{\pm}=a\pm\sqrt{a^2+2b}.
\end{equation}
In Eq. (24) $F[.]$ stands for the incomplete elliptic integral of the first kind \cite{22}. Equation (23) can be solved to write 
\begin{equation}
 f_g(\xi)=\frac{2}{\delta_-}\sqrt{\frac{b}{\delta_+}}sn(-i\sqrt{\frac{\delta_-}{2}}\xi,m).
\end{equation}
The results in Eq. (23) to Eq. (26) have been obtained by the use of Mathematica. It is of interest to note that for $m=1$ the Jacobi elliptic sine function $sn(.)$ in Eq. (26) reduces to hyperbolic tangent function \cite{22}. Under this constraint, in view of Eq. (22), the speed of the wave can be written in terms of parameters of the generalized mKdV equation (9) so that
\begin{equation}
 c=\pm i\sqrt{\frac{2}{3}\alpha\beta}.
\end{equation}
Since $c$ is real, Eq. (27) provides a restriction on the choice for the parameters $\alpha$ and $\beta$; one of them must be negative. For the positive value of $c$ we get from Eq. (26)
\begin{equation}
 f_g^{(+)}(\xi)=i\tan(\frac{\alpha\sqrt{i\sqrt{\frac{\beta}{\alpha}}}\xi}{6^{\frac{1}{4}}\sqrt{\beta}}).
\end{equation}
Similarly, for the negative value of $c$ we have
\begin{equation}
 f_g^{(-)}(\xi)=-i\tan(\frac{\alpha\sqrt{-i\sqrt{\frac{\beta}{\alpha}}}\xi}{6^{\frac{1}{4}}\sqrt{\beta}}).
\end{equation}
The superscripts $\pm$ on $f_g(\xi)$ have been used only to distinguish between the solutions of Eq. (21) belonging to the positive and negative values of $c$. For arbitrary choice for the values of $\alpha$ and $\beta$, say, $\alpha=-2$ and $\beta=2$ we have
\begin{equation}
 f_g^{(+)}(\xi)=\tanh(\frac{\xi}{6^{\frac{1}{4}}})\;\;\;\mbox{and}\;\;\;f_g^{(-)}(\xi)=i\tan(\frac{\xi}{6^{\frac{1}{6}}}).
\end{equation}
In the recent past the solutions similar to those in Eq. (30) for the generalized mKdV equation (9) were found \cite{23} by taking recourse to the use of first integral method \cite{24}. It was claimed that the solutions reported are new in the literature. But it appears that the such results for the  mKdV equation were obtained  by a number of authors \cite{25, 26, 27} during the first decade of the twenty first century. However, we point out that $f_g^{(+)}(\xi)$ is, in fact, a kink soliton solution of the generalized mKdV equation (9) since it goes to $\pm 1$ as $\xi\rightarrow \mp\infty$. Figure 4 gives the plots of $f_g^{(+)}(\xi)$ and $-f_g^{(+)}(\xi)$ as functions of $\xi$.
\begin{figure}[htb!]
\includegraphics[width=4.5cm]{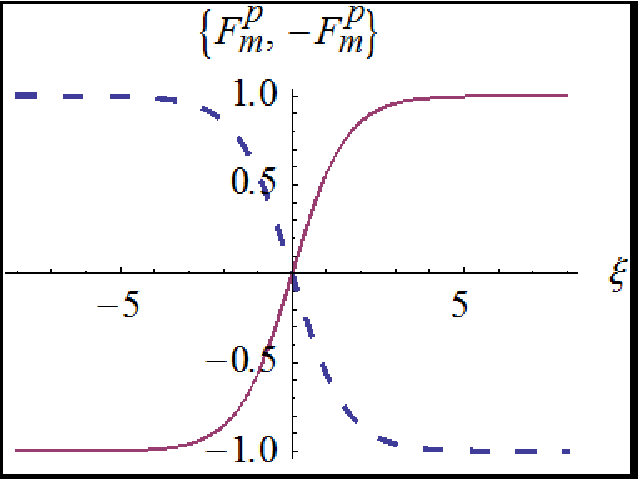}
\caption{The Kink and Anti-kink solitons for $\gamma=0.4$.}
\end{figure}
\begin{figure}[htb!]
\includegraphics[width=4.5cm]{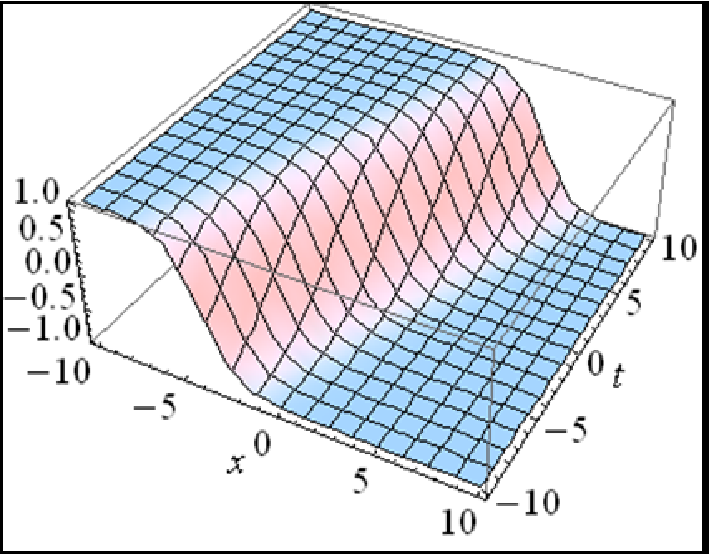}
\includegraphics[width=4.5cm]{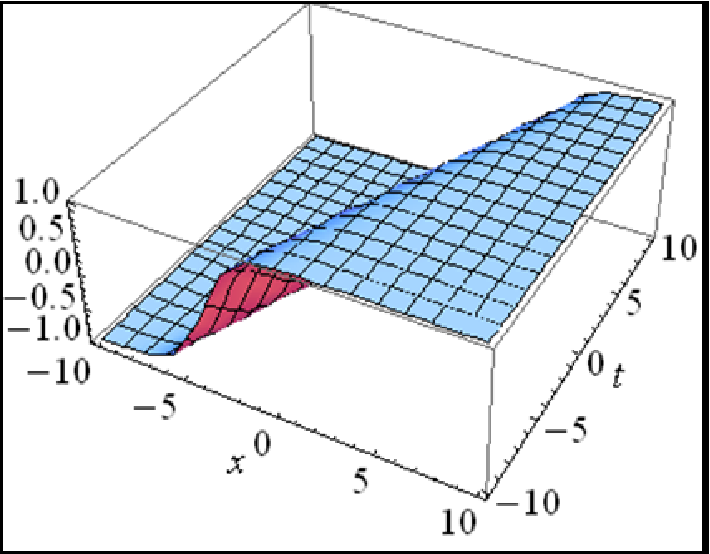}
\caption{(a) (upper)3D plot of the Kink soliton, (b) (lower) 3D plot of Anti-kink soliton.}
\end{figure}
The corresponding 3D plots are presented in Figs. 5(a) and 5(b) respectively.
\par In the above we obtained the travelling-wave solution of a generalized mKdV equation in terms of a Jacobi elliptic function and subsequently applied a constraint on it to find certain physically important soliton solutions. We shall now provide a similar treatment for the NLS equation (6) and try to derive some results for NLS solitons from travelling-wave solution of the equation. In order that it is appropriate to write
\begin{equation}
 u(x,t)=e^{i(\frac{cx}{2}+\eta t)}\nu(\xi),
\end{equation}
where $\eta$ is a well behaved real parameter. From Eq. (6) and Eq. (31) it is straightforward to construct the Duffing equation \cite{21}
\begin{equation}
 \nu''(\xi)-(\eta+\frac{c^2}{4})\nu(\xi)+\lambda\nu(\xi)^3=0.
\end{equation}
We have found solution of Eq. (32) in the form
\begin{equation}
 \nu(\xi)=2\sqrt{\frac{2}{\mu_-}}sn(2i\sqrt{\frac{\lambda}{\mu_+}\xi},m_n)
\end{equation}
with
\begin{equation}
 m_n=\frac{\mu_+}{\mu_-},\;\;\;\mu_\pm=c^2+4\eta\pm\sqrt{(c^2+4\eta)^2-32\lambda}.
\end{equation}
The expression for $\nu(\xi)$ in Eq. (33), the travelling wave of the NLS equation, is similar to that for the generalized mKdV equation, namely, $f_g(\xi)$ in Eq. (26). However, to obtain the NLS solitons from Eq. (33) we shall demand $m_n=1$. This gives the unknown parameter $\beta$ in terms of the parameter $\lambda$ of the NLS equation (6). In particular, we get two values of $\beta$ written as 
\begin{equation}
 \beta_{1,2}=-\frac{1}{4}(c^2\pm 4\sqrt{2\lambda}).
\end{equation}
For real values of $\lambda$, $\beta_1$ is always negative while $\beta_2$ is negative if $c^2>4\sqrt{2\lambda}$; otherwise $\beta_2$ is positive and we have  
\begin{subequations}
 \begin{equation}
 \nu_1(\xi)=i(\frac{2}{\lambda})^{\frac{1}{4}}\tanh((\frac{\lambda}{2})^{\frac{1}{4}}\xi)\;\;\;\mbox{for}\;\;\;\beta=\beta_1 
 \end{equation}
\mbox{and}
\begin{equation}
 \nu_2(\xi)=(\frac{2}{\lambda})^{\frac{1}{4}}\tanh(i(\frac{\lambda}{2})^{\frac{1}{4}}\xi)\;\;\;\mbox{for}\;\;\;\beta=\beta_2.
\end{equation}
\end{subequations}
In most studies on the NLS equation the parameter $\lambda$ is chosen as 2 such that $\nu_1(\xi)=i\tanh(\xi)$ and $\nu_2(\xi)=i\tan(\xi)$. It is interesting to note that these solutions are identical to the solutions in Eq. (30) for the generalized mKdV equation (9). This is, however, quite expected since the partial differential equations in the mKdV hierarchy can be viewed as a subsystem of the corresponding equations in the NLS hierarchy \cite{28}.
\par In addition to the kink and anti-kink soliton solutions of the mKdV and NLS equation, we now proceed to find
bell type soliton solutions of these equations. In the recent past Salas et al \cite{29} considered the third-order Duffing
equation
\begin{equation}
 q''(t)+\kappa_1q(t)++\kappa_3q(t)^3=0
\end{equation}
subject to initial condition $q(0)=q_0$ and $q'(0)=0$, and found its solution as
\begin{equation}
 q(t)=q_0cn(\sqrt{\kappa_1+\kappa_3q_0^2t},\frac{\kappa_3q_0^2}{2(\kappa_1+\kappa_3q_0^2)}).
\end{equation}
In close analogy with the result in Eq. (38), we have found a solution for the Duffing equation (21) corresponding to the mKdV equation to read
\begin{equation}
 f_g(\xi)=cn(\sqrt{b-a}\xi,\frac{b}{2(b-a)}).
\end{equation}
In writing Eq. (39) we used the initial conditions $f_g(0)=1$ and $f'_g(0)=1$. For $b/(2(b-a))=1$ or $b=2a$ the Jacobi elliptic cosine function goes over to the sec hyperbolic function so that we get from Eq. (39) the bell type soliton solution 
\begin{equation}
 f_g(\xi)=\sec h(\sqrt{\frac{c}{\beta}}\xi).
\end{equation}
of the mKdV equation.
\par We solved the Duffing equation (32) corresponding to the NLS equation for initial conditions $\nu(0)=1$ and $\nu'(0)=0$ in terms of Jacobi elliptic cosine function to write
\begin{equation}
 \nu(\xi)=cn(\xi\sqrt{\lambda-\beta-\frac{c^2}{4}},\frac{\lambda}{2(\lambda-\beta-\frac{c^2}{4})}).
\end{equation}
As in the case of mKdV equation the choice $\frac{\lambda}{2(\lambda-\beta-\frac{c^2}{4})}=1$ or $\beta=(2\lambda-c^2)$ gives
\begin{equation}
\nu(\xi)=\sec h(\sqrt{\frac{\lambda}{2}}\xi),
\end{equation}
the well known bell type NLS soliton. Comparing the results in Eq. (40) and Eq. (42) we see that the mKdV soliton is a function of its velocity while the NLS soliton is velocity independent.
\section*{4. Concluding remarks}
We have begun this article by noting the role of differential equations in studying time evolution of physical
systems and thus introduced their classification scheme. In the family of nonlinear partial differential equations
there exists a narrow window providing the so called integrable system, the equations of which can be solved by
analytic methods to obtain solitons. The solitons are, in fact, localized travelling waves, which preserve their shape
and velocity in interaction and have been found to appear in diverse settings ranging from electromagnetic waves
to matter waves in Bose-Einstein condensates \cite{30}. One remarkable property of integrable systems is that they
can effectively be linearized via the inverse spectral method. Currently, soliton theory is a large part of theoretical
and mathematical physics. It is of interest to note that solitons in water have been created even in the undergraduate laboratory \cite{31}.
\par We constructed different types of soliton solutions of the KdV, mKdV and NLS equations by going over to the travelling coordinate. The travelling waves of a generalized mKdV and NLS equations were found to satisfy the well known Duffing equation. In both cases solutions of the associated Duffing equations were used to obtain soliton solutions. In particular, solutions found in terms of $cn(u,m)$ led to bell type solitons while those found in terms of $sn(u,m)$ gave rise to kink
and anti-kink solitons.
\par Integrable equations that support soliton solutions are characterized by linear dispersive terms. Evolution equations with nonlinear dispersive terms are also of considerable physical interest. For example, the equation
\begin{equation}
u_t+(u^m)_x+u^n_{xxx}=0,\;\;u=u(x,t),\;\;m>1,\;\;1<n\leq 3
\end{equation}
was found by Rosenau and Hyman \cite{32} to account for pattern formation in liquids. It was shown that the solutions of Eq. (43) are free from the usual infinite tails of solitons and vanish identically outside a finite range. The solution of Eq. (43) was given the name compacton. The point at which two compactons collide is accompanied by the birth of a low-amplitude compacton-anticompacton pair. This represents a point of contrast with soliton-soliton interaction in integrable systems. It will be an interesting exercise to make use our travelling-wave approach to solve Eq. (43) for different values of
$m$ and $n$ with a view to examine the role of nonlinearity on the structure of compactons.
\vskip0.5cm
{\bf{Declaration}}
\newline The authors have no conflicts of interest to disclose.

\end{document}